\documentclass[pre,aps,twocolumn,showpacs,floatfix]{revtex4}
\usepackage{graphics}
\usepackage{epsfig}
\usepackage{amsmath}
\usepackage{amsthm, amssymb, color}

\usepackage{subfig}
\usepackage{sistyle}

\begin{document}

\newcommand{\red}[1]{{\color{red}#1}}

\newcommand{\eps}{\varepsilon}          
\newcommand{\vph}{\varphi}             
\newcommand{\vth}{\vartheta}          
\newcommand{\D}{{\rm d}}
\newcommand{\II}{{\rm i}} 
\newcommand{\arcosh}{{\rm arcosh\,}}  
\newcommand{\erf}{{\rm erf\,}}       
\newcommand{\wit}[1]{\widetilde{#1}} 
\newcommand{\wht}[1]{\widehat{#1}}  
\newcommand{\lap}[1]{\overline{#1}} 
\newcommand{\demi}{\frac{1}{2}}    
\newcommand{\rar}{\rightarrow}    
\newcommand{\gop}{\wht{\phi}_{\vec{0}}} 

\title{Ageing of the 2+1 dimensional Kardar-Parisi-Zhang model}

\author{G\'eza \'Odor (1), Jeffrey Kelling (2,3) and Sibylle Gemming (2,3)}

\affiliation{
(1) MTA TTK MFA Research Institute for Natural Sciences, \\
P.O.Box 49, H-1525 Budapest, Hungary \\
(2) Institute of Ion Beam Physics and Materials Research \\
Helmholtz-Zentrum Dresden-Rossendorf \\
P.O.Box 51 01 19, 01314 Dresden, Germany\\
(3) Institute of Physics, TU Chemnitz\\
09107 Chemnitz, Germany}

\begin{abstract}
Extended dynamical simulations have been performed on a 
2+1 dimensional driven dimer lattice gas model to estimate
ageing properties. The auto-correlation and the auto-response
functions are determined and the corresponding scaling exponents 
are tabulated. Since this model can be mapped onto the 2+1 
dimensional Kardar-Parisi-Zhang surface growth model, our results 
contribute to the understanding of the universality
class of that basic system. 
\end{abstract}
\maketitle

%%%%%%%%%%%%%%%%%%%%%%%%%%%%%%%%%%%%%%%%%%%%%%%%%%%%%%%%%%%%%%%%%%%%%%%%%%
\section{Introduction}
%%%%%%%%%%%%%%%%%%%%%%%%%%%%%%%%%%%%%%%%%%%%%%%%%%%%%%%%%%%%%%%%%%%%%%%%%%

Physical ageing occurring in different systems such as glasses, polymers,
reaction-diffusion systems or cross-linked networks has been studied 
in physics systematically \cite{St78}.
Ageing occurs naturally in irreversible systems, relaxing towards 
non-equilibrium stationary states (for a recent comprehensive
overview see \cite{HP}). In many systems a single dynamical
length scale $L(t)\sim t^{1/z}$ describes the dynamics out of 
equilibrium \cite{Bray94}, where $z$ is the dynamical exponent. 
In ageing systems the time-translation invariance is broken and they are 
best characterized by two-time quantities, such as the dynamical correlation 
and response functions \cite{Cug}. 
The dynamical scaling laws and exponents describing these 
functions characterize the non-equilibrium universality classes 
\cite{odorbook}. 

In the ageing regime: $s\gg \tau_{\rm m}$ and $t-s\gg \tau_{\rm m}$, 
where $\tau_{\rm m}$ is a microscopic time scale, one expects the
following laws for auto-correlation ($C(t,s)$) and auto-response 
($R(t,s)$) functions of the field $\phi$:
\begin{eqnarray}
C(t,s) &=& \left\langle \phi(t) \phi(s) \right\rangle
- \left\langle \phi(t)\right\rangle\left\langle \phi(s)\right\rangle
= s^{-b} f_C\left(\frac{t}{s}\right)
\label{1} \\
R(t,s) &=& \left. \frac{\delta \left\langle \phi(t)\right\rangle}{\delta
j(s)}\right|_{j=0}
= \left\langle \phi(t) \wit{\phi}(s) \right\rangle = s^{-1-a} f_R\left(\frac{t}{s}\right)
\nonumber
\end{eqnarray}
where $s$ denotes the start and $t>s$ the observation time,
$j$ is the external conjugate to $\phi$.
These laws include the so-called ageing exponents $a,b$ and the scaling 
functions, with the asymptotic behavior 
$f_{C,R}(t/s) \sim (t/s)^{-\lambda_{C,R}/z}$
and the auto-correlation and auto-response exponents $\lambda_{C,R}$.
In non-Markovian systems they can be independent, but symmetries can
relate them to each other via scaling laws (see \cite{HP,odorbook}).

The KPZ equation describes the evolution of a fundamental non-equilibrium
model and exhibits ageing behavior.
The state variable is the height function $h(\mathbf{x},t)$ in the 
$d$ dimensional space
\begin{equation}  \label{KPZ-e}
\partial_t h(\mathbf{x},t) = v + \nu\nabla^2 h(\mathbf{x},t) +
\lambda(\nabla h(\mathbf{x},t))^2 + \eta(\mathbf{x},t) \ .
\end{equation}
Here $v$ and $\lambda$ are the amplitudes of the mean and local growth
velocity, $\nu$ is a smoothing surface tension coefficient and $\eta$
roughens the surface by a zero-average, Gaussian noise field exhibiting
the variance
$\langle\eta(\mathbf{x},t)\eta(\mathbf{x^{\prime}},t^{\prime})\rangle = 
2 T \nu \delta^d (\mathbf{x-x^{\prime}})(t-t^{\prime})$.
The letter $T$ is related to the noise amplitude (the temperature in
the equilibrium system), $d$ is the spatial dimensionality of the system 
and $\langle\rangle$ denotes a distribution average.

Research on this nonlinear stochastic differential equation and
the universality class introduced by Kardar, Parisi and Zhang (KPZ) 
\cite{KPZeq} is in the forefront of interest nowadays again.
This is the consequence of emerging new techniques applied for the open
questions \cite{KCW,CorwinUof,Halp13,OK13,NCC13} and experimental 
realizations \cite{AFOR14}.
This equation was inspired in part by the stochastic
Burgers equation \cite{Burgers74} and can describe the dynamics of 
simple growth processes in the thermodynamic limit \cite{H90}, 
randomly stirred fluid \cite{forster77}, directed polymers in random media
(DPRM) \cite{kardar85}, dissipative transport \cite{beijeren85,janssen86},
and the magnetic flux lines in superconductors \cite{hwa92}.
In one dimension a mapping \cite{kpz-asepmap} onto the 
Asymmetric Exclusion Process (ASEP) \cite{Rost81} exists. 
In this case the equation is solvable due to the Galilean symmetry 
\cite{forster77} and an incidental fluctuation-dissipation symmetry 
\cite{kardar87}.

\begin{figure}[ht]
\begin{center}
\epsfxsize=70mm
\epsffile{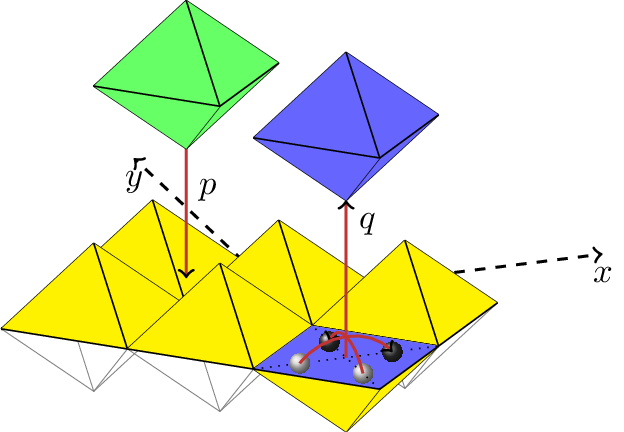}
\caption{(Color online) Mapping of the $2+1$ dimensional surface growth model
(octahedra) on the $2d$ particle model (bullets). Detachment (probability
$p$) and attachment (probability $q$) of octahedra correspond to Kawasaki
exchanges of pairs of particles along the bisectrix of the $x$ and $y$ axes. The
curved red arrows illustrate this as a superposition of two $1d$ processes.
The $2d$ square lattice to be updated is given by the crossing--points of the
dotted lines.
}
\label{fig:octahedron}
\end{center}
\end{figure}

It has been investigated by various analytical 
\cite{KCW,SE92,FT94,L95,F05,CCDW11} and numerical methods 
\cite{FT90,Halp12L,MPP,MPPR02,Reis05}, still there are several 
controversial issues.
Discretized versions of KPZ have been studied a lot in the
past decades \cite{meakin,barabasi,krug-rev}.
Recently we have shown \cite{asep2dcikk,asepddcikk} that 
the mapping between the KPZ surface growth and the ASEP \cite{kpz-asepmap} 
can straightforwardly be extended to higher dimensions.
In two dimensions the mapping is just the simple extension of the
rooftop model to the octahedron model as can be seen on
Fig.~\ref{fig:octahedron}.
The surface built up from octahedra can be described by the
edges meeting in the up/down middle vertexes. The up edges in the
$x$ or $y$ directions are represented by the slopes '$\sigma_{x/y} = 1$'-s, 
while the down ones by '$\sigma_{x/y} =-1$' in the model. 
This can also be understood as a special $2d$ cellular automaton, 
with the generalized Kawasaki updating rules
\begin{equation}\label{rule}
\left(
\begin{array}{cc}
   -1 & 1 \\
   -1 & 1 
\end{array}
\right)
 \overset{p}{\underset{q}{\rightleftharpoons }}
\left(
\begin{array}{cc}
   1 & -1 \\
   1 & -1 
\end{array}
\right)
\end{equation}
with probability $p$ for attachment and probability $q$ for
detachment. By the lattice gas representation with 
$n_{x/y} = (1-\sigma_{x/y})/2$ occupation variables it describes
the oriented migration of self-reconstructing dimers.  
We have confirmed that this mapping using the parametrization:
$\lambda = 2 p/(p+q)-1$ reproduces the one-point functions of the
continuum model \cite{asep2dcikk,asepddcikk}.

This kind of generalization of the ASEP model can be regarded
as the simplest candidate for studying KPZ in $d>1$: a
one-dimensional model of self-reconstructing $d$-mers on the
$d$-dimensional space.
Furthermore this lattice gas can be studied by very efficient
simulation methods.
Dynamic, bit-coded simulations were run on extremely large sized 
($L\times L$) lattice gas models \cite{asepddcikk,GPU2cikk} and 
the surface heights, reconstructed from the slopes
\begin{equation}\label{height}
h_{i,j} = \sum_{l=1}^i \sigma_x(l,1) + \sum_{k=1}^j \sigma_y(i,k)
\end{equation}
were shown to exhibit KPZ surface growth scaling in $d = 1 - 5$
dimensions.

While ageing in glassy systems follows a complex phenomenology \cite{Cug} 
the dynamic Renormalization Group (RG) analysis of KPZ presented in 
\cite{Ker} suggests that the one-scale dynamic scaling hypothesis 
is not spoiled for the KPZ universality class. This has been tested
by simulation studies and the present work strengthens this view further.

Recently, in $2+1$ dimensions Daquila and T\"auber \cite{DT11} have 
simulated the long-time behavior of the density-density auto-correlation 
function of driven lattice gases \cite{beijeren85} with particle exclusion 
and periodic boundary conditions in one to three spatial dimensions. 
In one dimension, their model is just the ASEP. 
They generalized this driven lattice gas model to higher dimensions 
by keeping the ASEP dynamics in one of the dimensions and performing 
unbiased random walk in the orthogonal dimension(s).
In two dimensions they reported: $\lambda_C/z = 1$ and $b=-1$.
We will show here that our generalization of ASEP model, which
exhibits the surface growth scaling of the $2+1$ dimensional KPZ
model provides different auto-correlation exponents.

Even more recently Henkel et al \cite{HDP} have determined the following
ageing exponents of the $1+1$ dimensional KPZ equation: 
$a=-1/3$, $b=-2/3$, $\lambda_C=\lambda_R=1$ and $z=3/2$.
They solved the discretized KPZ equation (\ref{KPZ-e}) in the strong 
coupling limit \cite{Newm97}, or else the Kim-Kosterlitz (KK) 
model \cite{Kim89}. The KK model uses a height variable 
$h_i(t)\in\mathbb{Z}$ attached to the sites of a chain with $L$
sites and subject to the constraints $|h_i(t)-h_{i\pm 1}(t)|=0,1$, 
at all sites $i$.

%%%%%%%%%%%%%%%%%%%%%%%%%%%%%%%%%%%%%%%%%%%%%%%%%%%%%%%%%%%%%%%%%%%%%%%%%%
\section{Bit-coded GPU algorithms}
%%%%%%%%%%%%%%%%%%%%%%%%%%%%%%%%%%%%%%%%%%%%%%%%%%%%%%%%%%%%%%%%%%%%%%%%%%

The height of each surface site is thoroughly determined by two slopes, along
the $x$ and $y$ axes respectively, whose absolute values are restricted to
unity. Thus at each site two bits of information are required, hence a chunk of
$4 \times 4$ sites is encoded in one 32-bit word.

\begin{figure}[h!t]
 \centering
 \subfloat[dead border]{
  % \deadborder{1}
  \epsfxsize=30mm
  \epsffile{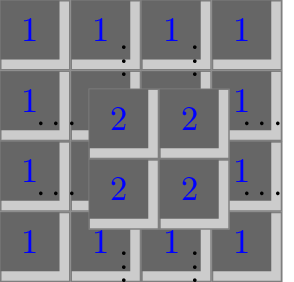}
  \label{fig:deadBorder}
  }
 \subfloat[double tiling]{
  % \doublecheckerboard{1}
  \epsfxsize=30mm
  \epsffile{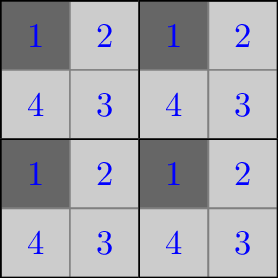}
  \label{fig:doubleTiling}
 }
 \caption{(Color online)
  Sketches of the domain decomposition methods used to parallelize the model.
  Regions that can be updated independently at a time are filled dark-grey.
  \protect\subref{fig:deadBorder}~Dead border scheme as used at device level.
  \protect\subref{fig:doubleTiling}~Double tiling scheme as used at work-group level.
}
\label{fig:domaindecomposition}
\end{figure}

Two different layers of parallelization are used that reflect the two
layered compute architecture provided by GPUs~\cite{NVCPG}: not communicating
blocks at \emph{device level} and communicating threads at 
\emph{work-group level}.
Parallelization of the algorithm is enabled by splitting the system into
spatial domains, which can be updated independently for a limited time without
introducing relevant errors.
At device layer a domain decomposition scheme using dead borders is
employed, see figure~\ref{fig:deadBorder}. Here conflicts at the
subsystem borders are avoided by not updating them.
A random translation is applied to the origin of
the decomposition periodically. These translations are restricted to multiples
of four sites, because $4\times4$ sites are encoded in one 32-bit word.
At work-group level a \emph{double tiling} decomposition is employed, see
figure~\ref{fig:doubleTiling}. Here the tiles assigned to different work-items
are split into $2^d$ domains.
In our two-dimensional problem, this creates $2^2$ sets of non-interacting
domains, each set consisting of one domain out of every tile. The active set of
domains is randomly chosen before each update.

For random number generation, each thread uses a 64-bit linear congruential
generator. The threads skip ahead in the sequence, in order to take numbers 
from disjunct sub-sequences.~\cite{weigel12}

A more detailed description of our CUDA implementation can be found
in~\cite{cpc11,EPJST12}. For this work we added the capability to perform 
simulations with arbitrary probabilities $p$ and $q$. 
Benchmarks, comparing our GPU implementation on a Tesla C2070 to the 
optimized sequential CPU implementation running on an Intel Xeon X5650 
at \SI{2.67}{GHz}, have shown a speedup-factor of about 230 for the raw 
simulation. The basic version from~\cite{EPJST12},
which contains less computational effort per update, reaches a raw simulation
speedup of about 100, in the same setup.

We decided to not implement space-dependent disorder in our GPU code, 
because we only need this code for the very small fraction of the computation 
before the waiting time~$s\leq\SI{100}{MCS}$. Thus the projected benefit 
regarding time-to-solution would not have justified the effort. 
Simulations to obtain auto-response calculations were performed using the 
CPU code up to the waiting time and then continued on
the GPU. The overall speedup-factor obtained by using a GPU in our use-case was
about 17. The difference to the number stated above results from using the CPU
code until reaching the waiting time and, predominantly, from the computation of
the auto-response not being done in parallel. In the auto-correlation runs
measurements where performed asynchronously with the simulation, in both CPU and
GPU versions. For these runs the gross speedup from using a GPU is about 50.

Applying any kind of domain decomposition to a stochastic cellular automaton
introduces an error. This error is kept small by keeping the ratio between the
volume of domains and the number of updated sites between synchronization event
large as well as by conserving the equidistribution of site-selection as best as
possible. The validity of the results was checked primarily by comparing with
results obtained with the sequential CPU implementation. For
the auto-correlation of slopes we noticed possible signs of saturation below
$C_n \lesssim \num{1e-4}$. This gives an upper limit for the accuracy of our 
GPU results, independent of statistics. 
Further investigations suggest, that the above-mentioned restriction of the 
translations of the decomposition-origin to multiples of four sites may be 
the sole source of this error. This restriction impairs the equidistribution 
of site-selection, while not enough to measurably change $W^2$ scaling, 
enough to visibly change the auto-correlation behavior of the system. 
We assume that this problem can be taken care of by removing this
restriction in the future.

%%%%%%%%%%%%%%%%%%%%%%%%%%%%%%%%%%%%%%%%%%%%%%%%%%%%%%%%%%%%%%%%%%%%%%%%%%
\section{Ageing simulations}
%%%%%%%%%%%%%%%%%%%%%%%%%%%%%%%%%%%%%%%%%%%%%%%%%%%%%%%%%%%%%%%%%%%%%%%%%%
 
We have run simulations for linear sizes: $L=2^{12}, 2^{13}, 2^{15}$ 
of independent samples $40000, 30000, 2000$ (respectively),
by starting from half filled (striped) lattice gases.
The time between measurements increases exponentially
\begin{equation}
t_{i+1} = (t_i  + 10) \cdot \mathrm{e}^m, \quad \text{with} \quad m > 0, \quad t_0 = 0,
\end{equation}
when the program calculates the heights $h_{\vec{r}}$ via Eq.~(\ref{height}) 
at each lattice site $\vec{r} = (i,j)$ and writes out the auto-correlation 
and the auto-response values to files, which are analyzed later. 
We used $s=30,100, 300$ start times in the two-point function measurements.
By simple scaling the morphology of the surface is characterized by the 
roughness
\begin{equation}
W^2(L,t) = \frac{1}{L^2} \sum_{\vec{r}}^{L^2} \left\langle \left(h_{\vec{r}}(t) 
- \overline{h}(t) \right)^2 \right\rangle
\end{equation}
on a lattice with $L^2$ sites and average height
{$\overline{h}(t) = L^{-2} \sum_{\vec{r}} h_{\vec{r}}(t)$},
which obeys the scaling relation
\begin{equation}
\hspace{-0.10truecm} W(L,t) = L^{\alpha} f\left(t L^{-z}\right)  , \;
f(u) \sim \left\{\begin{array}{ll} u^{\beta} & \mbox{\rm ;\ for $u\ll 1$} \\
                                \mbox{\rm const.}  & \mbox{\rm ;\ for $u\gg 1$} 
\end{array} \right.
\end{equation}
In this form $\beta$ is the growth exponent and the roughness exponent is
$\alpha=\beta z$. Throughout this paper we used the estimates
from our previous high precision simulation study \cite{GPU2cikk}: 
$\alpha=0.393(4)$, $\beta=0.2415(15)$ and the dynamical scaling 
exponent $z=\alpha/\beta=1.627(26)$.

Similarly to the one-dimensional case we considered here the two-time 
temporal correlator
\begin{eqnarray}
C(t,s) &=&
\left\langle \left( h(t;\vec{r}) - {\left\langle
\overline{h}(t;\vec{r})\right\rangle} \right)\left( h(s;\vec{r}) -
{\left\langle \overline{h};\vec{r}(s)\right\rangle} \right)
\right\rangle
\nonumber \\
&=& \left\langle h(t;\vec{r}) h(s;\vec{r}) \right\rangle -
{\left\langle \overline{h}(t;\vec{r})\right\rangle \left\langle
\overline{h}(s;\vec{r})\right\rangle}
\nonumber \\
&=& s^{-b} f_C\left( \frac{t}{s} \right) , \label{Ceq}
\end{eqnarray}
were $\langle\rangle$ denotes averaging over sites and independent runs.

The auto-correlation exponent can be read-off in the $(t/s)\to\infty$
limit: $f_C(t/s) \sim (t/s)^{-\lambda_C/z}$ and since 
$W^2(t;\infty)=C(t,t)=t^{-b} f_C(1)$ the $b=-2\beta$ relation holds.
The simulations were tested by blocking the communication in
one of the directions and comparing the results with those of the 
one dimensional KPZ ageing results~\cite{HDP}. As we found perfect
match we assume that our two-dimensional results give reliable numerical
estimates. 

\begin{figure}
\begin{center}
\epsfxsize=70mm
\epsffile{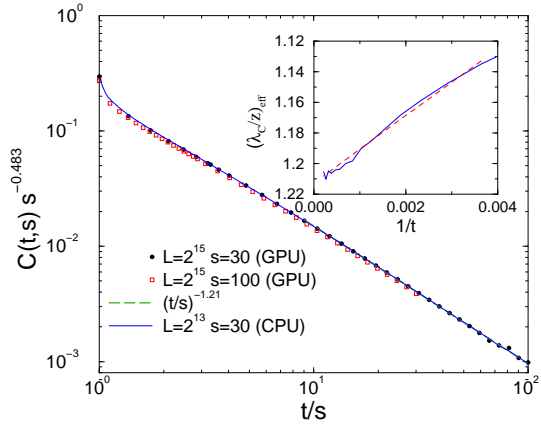}
\caption{(Color online) Auto-correlation function scaling of the
height variables for $L=2^{15}$, $s=30$ (bullets), $s=100$ 
(squares) and $L=2^{13}$, $s=30$ (line).
The dashed line shows a power-law fit for $t/s > 10$ with the slope
$-1.21$. Inset: Local slopes of the $L=2^{13}$, $s=30$ data defined
as (\ref{eff}). The dashed line shows a power-law fit.}
\label{slopes}
\end{center}
\end{figure}

As Fig.~\ref{slopes} shows we could obtain remarkable data 
collapse for $s=30$ and $s=100$ by simulating $2^{15}\times2^{15}$ 
sized systems on GPUs. Throughout this paper all quantities plotted 
are dimensionless. For smaller sizes it is more difficult to 
reach a regime of $t/s$ large, due to the low signal/noise ratio.
We performed careful correction to scaling analysis by calculating the 
local slopes of the auto-correlation function exponents for $t\to\infty$. 
The effective exponents can be estimated similarly as in case of other 
scaling laws \cite{odorbook} as the discretized, logarithmic derivative
\begin{equation}  \label{eff}
(\lambda/z)_{eff}(t_i) = \frac {\ln C(t_i) - \ln C(t_{i+1})} 
{\ln(t_{i+1}) - \ln(t_{i})} \ , 
\end{equation}
and we extrapolated to the asymptotic behavior with the form 
\begin{equation}  \label{slfit}
(\lambda/z)_{eff}(t_i) = \lambda/z + a t^x \ ,
\end{equation}
for $t > 250$. On the inset of Fig.~\ref{slopes} one can see
a roughly linear approach in $1/t \to 0$ with $\lambda_{C}/z = 1.21(1)$
and $a=20$. However, periodic corrections to scaling can also be observed,
which are the consequence of density fluctuations being transported 
through a finite system by kinematic waves
\cite{GMGB07,asepddcikk}.     

This provides $\lambda_{C}=1.97(3)$, in a marginal agreement with the 
$\lambda_C=d$ conjecture of \cite{KalKrug}, based on a purely geometric 
argument.
In \cite{Ker} a $2+1$ dimensional ballistic deposition model of 
linear size $L=240$ and $t\le 1000$ was simulated. 
Scaling with the form $C_L(t,s) \propto (t/s)^{-1.65(5)}$ is reported, 
which is out of the error margin of our large scale simulations 
and of the scaling law $\lambda_C/z = (d+4)/z -2 \simeq 1.08(5)$
derived in \cite{Ker}.

\begin{figure}
\begin{center}
\epsfxsize=70mm
\epsffile{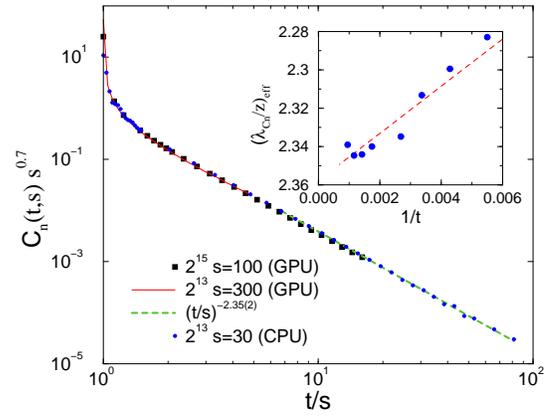}
\caption{(Color online) Auto-correlation function scaling of the
lattice gas variables for : $L=2^{15}$ $s=100$ (boxes, GPU), 
$L=2^{13}$, $s=300$ (line, GPU), $L=2^{13}$, $s=30$ (squares, CPU). 
The dashed line shows a power-law fit: $\sim (t/s)^{-2.35(2)}$ for $t/s >4$. 
Inset: Local slopes of the $L=2^{13}$, $s=30$ data defined
as (\ref{eff}). The dashed line shows a power-law fit. }
\label{acs}
\end{center}
\end{figure}

We have also calculated the auto-correlation of the density
variables  
\begin{eqnarray}
C_n(t,s) &=&
\left\langle \left( n(t;\vec{r}) - {\left\langle
\overline{n}(t;\vec{r})\right\rangle} \right)\left( n(s;\vec{r}) -
{\left\langle \overline{n};\vec{r}(s)\right\rangle} \right)
\right\rangle
\nonumber \\
&=& \left\langle n(t;\vec{r}) n(s;\vec{r}) \right\rangle -
{\left\langle \overline{n}(t;\vec{r})\right\rangle \left\langle
\overline{n}(s;\vec{r})\right\rangle}
\nonumber \\
&=& s^{-b'} f'_C\left( \frac{t}{s} \right) , \label{Cseq}
\end{eqnarray}
however, that decays much faster than the height auto-correlator and 
obtaining reasonable signal/noise ratio requires much higher statistics. 
This constrained the maximum time we could reach. Still, as Fig.~\ref{acs} 
shows, good data collapse could be achieved with $b'=-0.70(1)$ and 
$C_n(s,t)\propto (t/s)^{-2.35(2)}$ asymptotically.
In fact the height-height and the density-density correlation
functions can be related, since we have a one-dimensional motion of
dimers, for which \cite{PS02} derived
\begin{equation}
C_n(r,t) \sim \frac{\partial^2}{\partial r^2} C(r,t) \ .
\end{equation}
Indeed, a $2/z \simeq 1.23$ difference seems to connect the 
measured auto-correlator exponents $\lambda_C/z =1.21(1)$ and 
$\lambda_{C_n}/z = 2.35(2)$ fairly well.

Next, we investigated the scaling of the auto-response function
in a similar way as described in \cite{HDP}. Initially we applied
a space-dependent deposition rate $p_i=p_0 + a_i \eps /2$ with 
$\Delta=\pm 1$ and $\eps=0.005$ a small parameter. 
Then later on we used the same stochastic noise $\eta$ (random sequences), 
in two realizations. System A evolved, up to the waiting time $s$, with the 
site-dependent deposition rate $p_i$ and afterward, with the 
uniform deposition rate $p_0 = (1-q_0) = 0.98$.
System B evolved always with the uniform deposition rate $p_i=p_0$.
The time-integrated response function is
\begin{eqnarray}
\lefteqn{ \chi(t,s) = \int_0^s \!\!\D u\: R(t,u) }  \label{Req} \\
&=&
\frac{1}{L^2} \sum_{\vec{r}}^{L^2} \left\langle \frac{h_{\vec{r}}^{(A)}(t,s) -
h_{\vec{r}}^{(B)}(t)}{\eps \Delta}\right\rangle
= s^{-a} f_{\chi}\left( \frac{t}{s} \right)
\nonumber
\end{eqnarray}
Asymptotically for $(t/s)\to\infty$ one can read off the auto-response 
exponent: $f_{\chi}(y) \sim (t/s)^{-\lambda_R/z}$.
\begin{figure}
\begin{center}
\epsfxsize=70mm
\epsffile{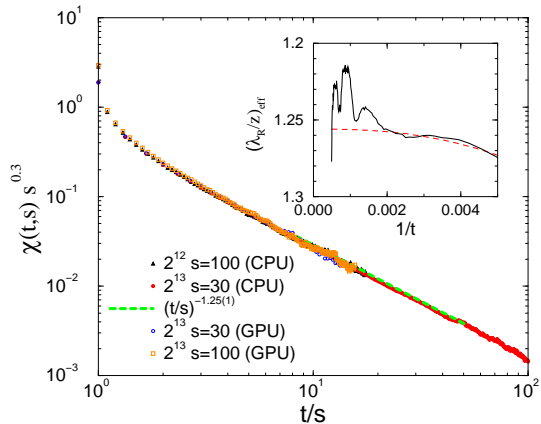}
\caption{(Color online)
Auto-response function scaling for $L=2^{12}$ (CPU) $s=100$ triangles) 
$s=30$ dots and $L=2^{13}$ (GPU) $s=30$ circles, $s=100$ squares.
The dashed line shows an asymptotic fit $\sim (t/s)^{1.25(1)}$ for
the  $L=2^{13}$, $s=30$ data in the $10 < t/s < 50$ region. 
Inset: Local slopes of the $L=2^{13}$, $s=30$ data defined
as (\ref{eff}). The dashed line shows a power-law fit. }
\label{alphas}
\end{center}
\end{figure}

Again, first we tested our programs by comparing the results against 
the one dimensional KPZ case \cite{HDP} by restricting the communication
among particles to one of the directions. 
Then we run large scale simulations on CPUs for $L=2^{13}$ up to $30000$
samples and for GPUs for $L=2^{13}$ up to $37000$ samples.
Hardware independence was confirmed and a good scaling collapse was achieved by the exponents shown on Fig.~\ref{alphas}.
We performed local slope analysis similarly as in case of the auto-correlations
(\ref{eff}). A least squares error power-law fitting (\ref{slfit}) resulted 
in a roughly quadratic approach to the asymptotics 
$\lambda_{R}/z = 1.255(10)  - 200 \ t^2$ as shown in the inset of 
Fig.~\ref{alphas}. Corrections to the long-time scaling 
are stronger and they suggest oscillating convergence as in the case 
of the auto-correlations.
Most obvious is that $\lambda_R\ne\lambda_C$, so the 
fluctuation-dissipation relation $TR(t,s)=-\partial_r^2 C(t,s)$, 
which is fulfilled in one dimension due to the time-reversal 
symmetry \cite{Deke75,forster77,CCDW11} is broken here.
The ageing exponents are different from the 1d KPZ \cite{HDP} and those
of the 2d driven lattice gas model of \cite{DT11}. They are summarized in
table~\ref{tab1}.

\begin{table}[h]
\begin{tabular}{|cccccc|} \hline
~$a$~   & ~$b$~  	& ~$\lambda_R$~ &	$\lambda_C$~ & ~$\beta$~ & ~$\alpha$~ \\ \hline
$0.30(1)$&  $-0.483(2)$	& $2.04(3)$     & 	$1.97(3)$    & $0.2415(15)$  & $0.393(4)$    \\
\hline
\end{tabular}
\caption{Scaling exponents of the $d=2+1$ dimensional KPZ class.\label{tab1}}
\end{table}

%%%%%%%%%%%%%%%%%%%%%%%%%%%%%%%%%%%%%%%%%%%%%%%%%%%%%%%%%%%%%%%%%%%%%%%%%%
\section{Conclusions and discussion}
%%%%%%%%%%%%%%%%%%%%%%%%%%%%%%%%%%%%%%%%%%%%%%%%%%%%%%%%%%%%%%%%%%%%%%%%%%

We have extended our previous, bit-coded 2d driven dimer lattice gas model
simulations with auto-correlation and auto-response measurement capability
in order to investigate the ageing behavior. This gas can be mapped onto
a surface growth (octahedron) model, which exhibits KPZ surface scaling
exponents, thus our height auto-correlation and auto-response functions
describe the ageing properties of two dimensional KPZ surfaces. By performing
extensive simulations both on CPUs and GPUs we have determined the ageing
exponents for this universality class. 
The auto-correlation exponents are different from those of the 
two-dimensional driven lattice gas \cite{DT11} and of the simulations
of \cite{Ker}, however fairly good agreement was found
with the hypothesis of \cite{KalKrug}.
Weak violation of the fluctuation-dissipation relation is confirmed 
numerically. 

We have also provided numerical estimates for the auto-correlation exponents 
for the density variables of the dimer lattice gas.
In one-dimensional models of self-reconstructing $d$-mers conservation laws 
resulted in initial condition dependent sectors, with different power-laws 
\cite{M97,BGS07}, placing a question mark on the universality.
In higher dimensions exclusion effects are less relevant \cite{odorbook},
furthermore, due to the KPZ surface mapping, not all initial conditions
and particle configurations are allowed. Still a more detailed study in
this direction would be very interesting.
 
The performance of the GPU code with respect to the CPU algorithm is higher 
by about a factor of 230.
Our method is capable to test numerically predictions of the Local Scale
Invariance hypothesis (see \cite{HP}) and is straightforwardly extensible
to higher dimensions \cite{asepddcikk}. For $p=q$ in the octahedral 
adsorption-desorption model the long-time dynamics is governed by 
the Edwards-Wilkinson scaling \cite{EW,asep2dcikk}. Numerical test
of the ageing properties with respect to analytical results is
planned in a future work.

Following the submission of this paper we learned that Tim Halpin-Healy
obtained auto-correlation results for different other models: 
Restricted Solid on Solid, KPZ Euler, DPRM belonging to the KPZ class
(for definitions see \cite{Halp12L}), which agree with ours provided 
an overall, model dependent multiplication factor is applied \cite{unpub}.

\vskip 1.0cm

\noindent
{\bf Acknowledgments:}\\

Support from the Hungarian research fund OTKA (Grant No. K109577),
and the OSIRIS FP7 is acknowledged. 
We thank Karl-Heinz Heinig for initiating the German-Hungarian
cooperation and laying the ground for our large scale simulations, 
Uwe T\"auber, Tim Halpin-Healy and Joachim Krug for their useful comments.
The authors thank NVIDIA for supporting the project with high-performance
graphics cards within the framework of Professor Partnership.
Jeffrey Kelling thanks Karl-Heinz Heinig for his mentoring on bit-coded
simulations and Bartosz Liedke for additional support in this matter.

\end{document}